\documentstyle[preprint,pra,aps]{revtex}
\tightenlines
\begin{document}
\draft
\title{Polarizabilities and parity non-conservation in the Cs atom and
limits on the deviation from the standard electroweak model}

\author{ V. A. Dzuba$^*$, V. V. Flambaum, O. P. Sushkov }
\address{School of Physics, University of New South Wales,
	Sydney, 2052, Australia}
\date{\today}
\maketitle
\begin{abstract}
A semi-empirical calculation of the 6s - 7s Stark amplitude $\alpha$ 
in Cs has been performed using the most accurate measurements and
calculations of the electromagnetic amplitudes available. This is then 
used to extract the parameters of the electroweak theory from experimental
data. The results are: $\alpha = 269.0 (1.3) a_0^3$, weak charge of Cs 
$Q_W = -72.41(25)_{exp} (80)_{theor}$, deviation from the Standard
model $S = -1.0(.3)_{exp} (1.0)_{theor}$ and limit on the mass of the 
extra Z-boson in $SO(10)$ model $M_{Z_x} > 550 GeV$.
\end{abstract}



Experiments suggested in \cite{Bouchiat74} for measuring parity 
nonconservation (PNC) in heavy atoms have provided an important confirmation
\cite{Barkov,Edwards95,Wood97} 
of the standard model of elementary particles. Combining the 
recent very accurate measurements of parity nonconservation in the Cs atom
\cite{Wood97} with theoretical calculations 
\cite{Dzuba89a,Blundell90} gives one a possibility to study
new physics beyond the standard model.
The measured nuclear spin-independent part of the PNC effect in Cs 
\cite{Wood97} is of the form (we use the analysis from \cite{Flambaum97})
\begin{eqnarray}
	- \frac{Im(E_{PNC})}{\beta} = 1.5939(56) \frac{mV}{cm}
\label{PNCe}
\end{eqnarray}
where $E_{PNC}$ is the PNC E1-amplitude of the 6s - 7s transition and
$\beta$ is the vector polarizability of the transition.
The theoretical values of $E_{PNC}$ are as follows:
\begin{eqnarray}\label{PNCt}
	E_{PNC} = -i|e|a_0 10^{-11} (-\frac{Q_W}{N}) \left\{ \begin{array}{ll}
                   0.908(10) & \mbox{Ref} \cite{Dzuba89a} \\
                   0.905(9)  & \mbox{Ref} \cite{Blundell90}\\
\end{array} \right.
\end{eqnarray}
Here $Q_W$ is the weak charge of the cesium nucleus and $N$ is the number of 
neutrons.

The method for {\it ab initio} calculations of $E_{PNC}$ that we used in
\cite{Dzuba89a} was based on an all-orders summation of the dominating 
diagrams of the many-body perturbation theory in the residual Coulomb 
interaction using a relativistic Hartree-Fock basis set and Green's 
functions. This technique has been described in 
\cite{Dzuba89a,Dzuba87a}.

We took into account direct and exchange polarization of the atomic core by
the external electric field and the weak nuclear potential using the
Time-Dependent Hartree-Fock method
(summation of the ``RPA with exchange''  chain of diagrams), calculated 
second-order correlation corrections and 3 series of dominating 
higher-order diagrams: 
\begin{enumerate}
\item Screening of the electron-electron interaction. This is a collective
phenomenon and so the corresponding chain of diagrams is enhanced by a
factor approximately equal to the number of electrons in the
external closed subshell (the 5p electrons in Cs).
We  stress that our approach takes into account screening
diagrams with double, triple and higher core electron excitations 
\cite{comment}, in contrast to popular pair equations (coupled cluster) 
method, where only double excitations were considered.
\item Hole-particle interaction. This effect is enhanced by the
large zero-multipolarity diagonal matrix elements of the Coulomb 
interaction.
\item Iterations of the self-energy operator ( ``correlation potential'').
This chain of diagrams describes the nonlinear effects of the correlation 
potential and is enhanced by the small denominator, which is the energy
for the excitation of an external electron (in comparison with the
excitation energy of a core electron).
\end{enumerate}

The error in the theoretical value was tested in many different ways:
by estimating the contribution of the unaccounted higher-order diagrams,
by comparing  the calculated and measured values of the energy levels, 
the fine and hyperfine structure intervals, the probabilities of 
electromagnetic transitions, etc. (see Ref. \cite{Dzuba89a}).
The result for the PNC amplitude practically did not change when we
introduced  factors into the correlation potential to fit the
energy levels (in imitation of the unaccounted higher-order diagrams).
Important tests of our method included predictions of the spectrum
\cite{Dzuba83} and electromagnetic transition amplitudes for the Fr
atom \cite{Dzuba95}, which is an analogue of Cs. Recently the positions of
many energy levels \cite{Liberman78} and some transition rates 
\cite{Orozco97} of Fr  were 
measured and found to be in excellent agreement with our predictions.

Our calculations of PNC for atoms with electron structures more complex 
than those of the alkaline atoms were proved to be accurate as well.
In a series of works done about ten years ago we claimed an accuracy of 3\%
for Tl \cite{Dzuba87b}, 8\% for Pb and 11\% for Bi \cite{Dzuba88}.
All these PNC effects were recently measured to an accuracy of about 1\%
\cite{Edwards95} and found to be in good agreement with 
our predictions. This
means that our estimates for the theoretical accuracy were correct and
probably even too pessimistic. For example, in our first calculation of
the Fr energy levels \cite{Dzuba83} we claimed the accuracy of our 
predictions to be about 0.5\% while the actual agreement with latter
measurements was found to be 0.1\%. The situation was similar for the
electromagnetic transitions 6s-6p$_{1/2}$ and 6s-6p$_{3/2}$ in Cs (see
below).
These numerous tests give us firm ground to believe that the theoretical
error in  $E_{PNC}$ (\ref{PNCt}) indeed does not exceed 1\%.

As can be seen from (\ref{PNCe}) an accurate value of the vector
transition polarizability $\beta$ is also required for the interpretation 
of the PNC measurements. 
There are no direct experimental measurements of $\beta$ and so the value
 $\beta = 27.0(2) a_0^3$ calculated in \cite{Johnson92}
was used for the interpretation of the PNC measurements. The theoretical
ratio of the  scalar transition polarizability $\alpha = -268(3) a_0^3$
to $\beta$ $(\alpha / \beta)_{theory} = -9.93(14)$ \cite{Johnson92} 
was in good agreement with the corresponding experimental value
 $(\alpha / \beta)_{exper} = -9.9(1)$ \cite{Hoff81} available at that time.
Since then the ratio  $(\alpha / \beta)$ was remeasured to a very high
accuracy:  $(\alpha / \beta) = -9.905(11)$ \cite{Cho97}. There have also
been new, very precise measurements of the lifetimes of the $6p_{1/2}$ and
$6p_{3/2}$ states of Cs \cite{Tanner92}. This allows us
to improve the accuracy in the determination of $\beta$, and thus in the
interpretation of the PNC measurements, by incorporating the new 
experimental results into our calculations.

The calculations were done using direct summation over the exact 
intermediate states
\begin{eqnarray}\label{beta}
	\beta = \frac{e^2}{9} \sum_n [ \langle 7s | r | np_{1/2} \rangle
	\langle np_{1/2} | r | 6s \rangle (\frac{1}{E_{7s} - E_{np_{1/2}}} -
	\frac{1}{E_{6s} - E_{np_{1/2}}})  \\
	-  \langle 7s | r | np_{3/2} \rangle
	\langle np_{3/2} | r | 6s \rangle (\frac{1}{E_{7s} - E_{np_{3/2}}} -
	\frac{1}{E_{6s} - E_{np_{3/2}}}) ] \nonumber
\end{eqnarray}
Here $ \langle s | r | np \rangle$ is an effective radial integral
for electromagnetic transitions between exact atomic eigenstates,
which are related to the reduced matrix elements by
\begin{eqnarray}
 \langle s || r || p_{1/2} \rangle =  \langle p_{1/2} || r || s \rangle =
 \sqrt{\frac{2}{3}} \langle s | r | p_{1/2} \rangle
\end{eqnarray}
\begin{eqnarray}
 \langle s || r || p_{3/2} \rangle =  - \langle p_{3/2} || r || s \rangle =
  \sqrt{\frac{4}{3}} \langle s | r | p_{3/2} \rangle
\end{eqnarray}
It easy to see that $\beta$ vanishes in the absence of the spin-orbit
interaction, which splits energy levels and radial integrals.
Thus, it is practically impossible to do accurate calculations of $\beta$
using experimental results due to the strong cancelation between different 
terms, which causes the relative statistical error to be larger. Therefore,
we calculated the scalar transition polarizability $\alpha$ instead and
used the measured ratio $\alpha / \beta$ to find $\beta$. Note however
that the calculation of $\alpha/\beta$ using theoretical radial integrals 
and experimental energies reproduces the experimental value for this ratio
to an accuracy of about 1\%.

The expression for $\alpha$ is given by

\begin{eqnarray}\label{alpha}
	\alpha = \frac{e^2}{9} \sum_n [ \langle 7s | r | np_{1/2} \rangle
	\langle np_{1/2} | r | 6s \rangle (\frac{1}{E_{7s} - E_{np_{1/2}}} +
	\frac{1}{E_{6s} - E_{np_{1/2}}})  \\
	+  2 \langle 7s | r | np_{3/2} \rangle
	\langle np_{3/2} | r | 6s \rangle (\frac{1}{E_{7s} - E_{np_{3/2}}} +
	\frac{1}{E_{6s} - E_{np_{3/2}}}) ] \nonumber
\end{eqnarray}
Here all of the major terms produce positive contributions. 
This reduces the error in the final result. 
98\% of the value of $\alpha$ is given by 
the intermediate 6p and 7p states. The 6p state
practically does not contribute to the error in the final result. Our
calculations of the 6s - 6p electromagnetic amplitudes were recently
confirmed to an accuracy of about 0.1\% by very accurate
experimental measurements \cite{Tanner92}. The 6p - 7s amplitudes
are also known from \cite{Bouchiat84} with an accuracy of  0.5\% and they 
agree with the theory.

The main source of error is the contribution of the 7p intermediate 
state. The radial integrals $\langle 6s | r| 7p_{1/2} \rangle$ and
 $\langle 6s | r| 7p_{3/2} \rangle$ are anomalously small
due to cancelations between different areas of the integration in the
single-particle amplitudes. These cancelations substantially increase 
the relative error in the calculated results. Because of this we use
the experimental values of the 6s - 7p transition amplitudes,
which have an accuracy of about 0.7\% \cite{Shab}. In \cite{Shab}
the relative oscillator strengths were measured using the lifetime of
the 6p$_{1/2}$ state measured in \cite{Link66} as a normalization point.
Recent measurements of the lifetime are more accurate \cite{Tanner92}.
Therefore we rescaled the experimental 6s - 7p amplitudes from 
\cite{Shab} using the new normalization. Note that the difference between 
$\langle 6s|r|7p_{1/2}\rangle$ and $\langle 6s|r|7p_{3/2}\rangle$ 
can be calculated very accurately. This is because it is proportional 
to the mixing between 7p and 6p states by the spin-orbit interaction. 
Indeed, perturbation theory in the spin-orbit interaction $\xi$ gives
\begin{eqnarray}\label{xi}
	\langle 6s|r|7p_{1/2}\rangle - \langle 6s|r|7p_{3/2}\rangle
	\sim \frac{\xi_{7p6p}}{E_{7p}-E_{6p}} \langle 6s|r|6p \rangle
	+ \cdots
\end{eqnarray}
The values of the energy levels and spin-orbit splitting can be reproduced 
almost exactly in the numerical calculations by introducing factors into
the correlation potential $\Sigma$ (since the accuracy of the {\it ab
initio} calculations is high, these factors are close to 1 anyway).
The calculated matrix element $\langle 6s|r|6p \rangle$ practically
coincides with the value obtained from the accurate measurements
of Ref. \cite{Tanner92}. Therefore, we believe that the absolute accuracy 
in the calculation of the difference between the doublet radial integrals
is always higher than the experimental accuracy (to avoid confusion we 
should note that we use Dirac wave functions, i.e. we do not expand in 
$\xi$ while doing calculations). 
Thus we can take the experimental value of $\langle 6s|r|7p_{1/2}
 \rangle$ which is measured more accurately, and find 
 $\langle 6s|r|7p_{3/2} \rangle$ using the calculated difference
 $\langle 6s|r|7p_{3/2} \rangle - \langle 6s|r|7p_{1/2} \rangle$.
Surprisingly, the result of this procedure gives precisely the result
of the measurement of the  $\langle 6s|r|7p_{3/2} \rangle$ amplitude,
which formally has a larger error (1.8\%). The ratio 
 $\langle 6s|r|7p_{3/2} \rangle / \langle 6s|r|7p_{1/2} \rangle$
also has a smaller experimental error (0.4\%) than the error in
 $\langle 6s|r|7p_{3/2} \rangle$ \cite{Shab}. Therefore, we may
assume that the actual relative error in the  $\langle 6s|r|7p_{3/2} \rangle$
is  0.7\%, similar to that in  $\langle 6s|r|7p_{1/2} \rangle$.
We use theoretical values of the  $\langle 7s|r|7p \rangle$
transition amplitudes since we believe that the expected theoretical
error here ( 0.3\%) is smaller than the experimental error. All higher
transitions, including continuum and core electron transitions, were also
calculated theoretically, even though their contribution was small
(see below).

The result of the calculation of $\alpha$ is as follows
\begin{eqnarray}\label{alpha-res}
	\alpha = \alpha(6p_{1/2}) + \alpha(6p_{3/2}) + \alpha(7p_{1/2})
	+ \alpha(7p_{3/2}) + \alpha(others) = \\
	-32.39(0.17) - 92.56(0.46) - (37.79 + 103.01)(1.14) -
	 3.25(0.20) = -269.0(1.3)
	\nonumber
\end{eqnarray}
We used experimental energy levels from \cite{Moore} and radial integrals 
from Table ~\ref{tab} to calculate the contributions of the 
6p and 7p states. We used both experimental and theoretical data
to select the ``best values'' of these integrals. Note that the
errors in $\alpha(7p_{1/2})$ and  $\alpha(7p_{3/2})$ are proportional
and so we added them.
When new data for electromagnetic amplitudes are available it
will be easy to refine this result by multiplying the corresponding
term by the ratio of the new amplitude to the old one.

This value of $\alpha$ combined with the measurements of $\alpha/\beta$
\cite{Cho97} gives 
\begin{eqnarray}\label{beta-res}
	\beta = 27.15(13) a_0^3 .
\end{eqnarray}
The result of the direct calculation using radial integrals from table
\ref{tab} is $\beta = 27.00$. The results of other works are
 $\beta = 27.0(2)$ \cite{Johnson92}, $\beta = 27.2(4)$ \cite{Bouchiat86b},
 $\beta = 27.3(4)$ \cite{Gilbert86}, $\beta = 27.17(35)$ \cite{Guena88}.
Using (\ref{beta-res}), the measurement (\ref{PNCe}), 
the mean value of the theoretical 
amplitudes (\ref{PNCt}) and $|e|/a_0^2 = 5.1422 \times 10^{12} mV/cm$,
we obtain
\begin{eqnarray}
	Q_W(exper) = -72.41(25)_{exper}(80)_{theor}.
\end{eqnarray}
Comparing this result for $Q_W$ with the theoretical value \cite{Marciano90}
\begin{eqnarray}\label{QW}
	Q_W(theor) = -73.20(13) - 0.8 S -0.005 T,
\end{eqnarray}
we can find the Peskin-Takeuchi parameter $S$ characterizing new physics
beyond the Standard model
(i.e. weak isospin conserving radiative corrections produced by new particles)
\begin{eqnarray}\label{S}
	S + 0.006 T =	-1.0(0.3)_{exper}(1.0)_{theor}.
\end{eqnarray}
We can also use the calculation of the extra $Z_x$-boson contribution in 
the $SO(10)$ model \cite{Marciano90}
\begin{eqnarray}
	\Delta Q_W = 0.4 (2N+Z)(\frac{M_W}{M_{Z_x}})^2 =
	 84.4(\frac{M_W}{M_{Z_x}})^2
\end{eqnarray}
to find the limit for the mass of this boson
\begin{eqnarray}\label{Mzx}
	M_{Z_x} > 550 GeV
\end{eqnarray}
The natural question is: can we refine the value of the $E_{PNC}$
calculation using experimental $E1$-amplitudes? Unfortunately, 
the experimental accuracy at the moment is not good enough to give 
an improvement.
For example, we can use the results of the work \cite{Johnson92},
where the direct sum-over-states approach was discussed in detail.
The theoretical result of the direct summation was
\begin{eqnarray}\label{PNCJ}
	E_{PNC} = -0.907(9) 10^{-11} i|e|a_0  (-\frac{Q_W}{N})
\end{eqnarray}
Replacing the $E1$-amplitudes calculated in \cite{Johnson92} 
(see table IV of that work) with the values from Table \ref{tab}
gives
\begin{eqnarray}
	E_{PNC} = -0.902(11)_{E1}(\sim 7)_{other} 10^{-11} i|e|a_0  
	(-\frac{Q_W}{N})
\end{eqnarray}
Here we separated the error coming from the 6p and 7p $E1$-amplitudes from
the error coming from all other sources, including the weak matrix elements
and the amplitudes for transitions to the states above 7p.
The error in the weak matrix elements can be roughly estimated using 
the deviation of the calculated hyperfine intervals from the experimental
values since both the weak and hyperfine interactions are approximately 
proportional to the density of the electron wave function near the nucleus.
Note that, the error from the $E1$-amplitudes exceeds the error in the 
theoretical values for the $E_{PNC}$ (\ref{PNCt}). To avoid confusion
we should stress that the calculation in Ref. \cite{Dzuba89a} was
based on the Green's function technique and does not contain partial
cancelations of the different terms which increase the error in the
direct sum-over-state approach.

In conclusion we would like to stress that accurate measurements of the 
$E1$-amplitudes (Table \ref{tab}) are very desirable for an improvement 
of the interpretation of the PNC measurements in Cs. For $\alpha$ the
most important improvement would be a more accurate value of the 
6s - 7p amplitude. An improvement for the 7s - 7p amplitude is also
very important because of the disagreement between theory and existing data.

The authors are grateful to David DeMille for helpful comments.
This work was supported by the Australian Research Council.

\begin{table}
\caption{Radial integrals used in the calculation of $\alpha$}
\label{tab}
\begin{tabular}{|ccc|}
 $np$ & $\langle 6s| r | np \rangle$ & $\langle np| r | 7s \rangle$~~ \\
\hline
 $6p_{1/2}$                     & -5.5091(75) & 5.190(27)~~ \\
 $6p_{3/2}$                     & -5.4824(62) & 5.605(27)~~ \\
 ~~$\Delta (6p_{3/2}-6p_{1/2})$ &  0.0267     & 0.4154 \\
 $7p_{1/2}$                     & -0.3460(26)  & -12.597(38)~~  \\
 $7p_{3/2}$                     & -0.5040(38)  & -12.372(37)~~  \\
 ~~$\Delta (7p_{3/2}-7p_{1/2})$ & -0.158      & 0.225 \\
\end{tabular}
\end{table}

\begin{thebibliography}{25}
\bibitem{comment}
	The point is that we exploit the Feynman diagram technique which
	contains all possible ``time ordering'' of the loops and therefore 
	screening diagrams contain any number of excited electrons.
\end{thebibliography}
\end{document}